\documentclass[aps,prb,twocolumn]{revtex4}  
\usepackage{epsfig}
\usepackage{amsmath}
\newcommand{\bpi}{\mbox{\boldmath$\pi$}}

\begin{document}
\title{Degeneracy and Strong Fluctuation-Induced
First-Order Phase Transition in the Dipolar Pyrochlore Antiferromagnet}
\author{O. C\'epas$^1$, A. P. Young$^2$ and B. S. Shastry$^2$}
\affiliation{
$1$  Institut Laue Langevin, BP 156, F-38042 Grenoble Cedex
9, France. \\
$2$ Department of Physics, University of California, Santa Cruz, 95064.}
\date{\today}

\begin{abstract}
We show that a continuous set of degenerate critical soft modes
strongly enhances the
first-order character of a fluctuation-induced first-order transition in the
pyrochlore dipolar Heisenberg
antiferromagnet. Such a degeneracy seems essential to
explain the strong first-order transition recently observed in
Gd$_2$Sn$_2$O$_7$. We present some evidence from Monte-Carlo simulations and a
perturbative renormalization group expansion.
\end{abstract}

\pacs{PACS numbers:} \maketitle
\section{Introduction}

In this paper we study phase transitions in Heisenberg magnets on the
pyrochlore lattice, which consists of corner sharing
tetrahedra\cite{Reimers,Moessner}.  The motivation is partly
theoretical, to understand the behavior of highly frustrated systems
when there is a degeneracy, or near degeneracy, between different
ordered states. There is also experimental motivation since
experiments on Gd$_2$Sn$_2$O$_7$ and Gd$_2$Ti$_2$O$_7$ have shown a
rich behavior,\cite{Raju,Bonville} including multiple field
transitions\cite{Ramirez,Petrenko} which we would like to understand.

For the family of rare earth pyrochlore systems it is well
known\cite{Raju} that dipole-dipole interactions are important, since
the angular momentum is large ($S=7/2$ for Gd).  If one adds nearest
neighbor exchange to dipole-dipole interactions, the Fourier transform
of the total interaction $J(\mathbf{q})$ is virtually
independent\cite{notedeg} of $\mathbf{q}$ (and takes its minimum
value) along the (1,1,1) directions of the reciprocal space.\cite{Raju,Palmer} This means that
the magnetic ordering wave-vector could, potentially lie anywhere along
these lines.

\begin{figure}[htbp]
\centerline{
\psfig{file=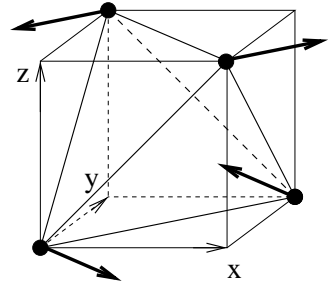,width=5cm,angle=-0}}
\vspace{-1cm}
\caption{The A state ($\bf{q}=\bf{0}$) of the pyrochlore lattice
stabilized at low temperatures by the dipole-dipole interactions
(Refs. [\onlinecite{Raju,Palmer,Ramirez,Cepas}]). Since the ordering
is at $\mathbf{q} = 0$, all tetrahedra have the same spin
configuration as the one shown.  In the figure, all the spins lie onto
the $(xy)$ plane and form pairs of antiparallel spins that are
parallel to the opposite edge of the tetrahedron they belong to. There
are equivalent $(xz)$ and $(yz)$ states. The magnetic order is
therefore characterized by a $n=3$-component order parameter, $\psi$ with
$(xy)$ corresponding to $\psi=(1,0,0)$ at T=0.}
\label{Astate}
\end{figure}

It turns out that the phases at the endpoints,
$\mathbf{q}=\mathbf{0}$ (denoted, following Ref.~\onlinecite{Cepas}, by A)
and $\mathbf{q}=(\pi, \pi, \pi)$ (denoted by $\bpi$ or B) are particularly
important. The ordering expected at $\mathbf{q}=\mathbf{0}$ (A-type)
has been discussed
by several authors\cite{Raju,Palmer,Ramirez,Cepas} and is shown in
Fig.~\ref{Astate}. A possible ordering at  $\mathbf{q}= \bpi$ (B-type) has
been proposed in Ref. [\onlinecite{Cepas}].

Although one would imagine that Gd$_2$Sn$_2$O$_7$ and
Gd$_2$Ti$_2$O$_7$ should be quite similar, since the crystal
structures are the same (apart from a very small difference in the
lattice constant), it is found that Gd$_2$Sn$_2$O$_7$ has a
\textit{strong} first-order transition,\cite{Bonville} while
Gd$_2$Ti$_2$O$_7$ has a second-order transition.\cite{Champion}
Furthermore, Gd$_2$Ti$_2$O$_7$ orders at
$\mathbf{q}=\mbox{\boldmath$\pi$}$\cite{Champion,Stewart} while
Gd$_2$Sn$_2$O$_7$ appears more compatible with the A
phase.\cite{Bertin} While the small change in lattice parameter
between Gd$_2$Sn$_2$O$_7$ and Gd$_2$Ti$_2$O$_7$ could change somewhat
the exchange constants, it seems remarkable that the nature of the
ordering changes so dramatically. We would like to understand such a
delicate dependence of ordering on exchange constants.

The first order nature of the transition observed in Gd$_2$Sn$_2$O$_7$
is at variance with mean-field theory which predicts\cite{Cepas} a
second-order transition. In order to clarify the order of the
transition, and to see whether it is affected by longer range exchange
interactions which lift the degeneracy of $J(\mathbf{q})$, we have
performed Monte Carlo simulations of the classical dipolar Heisenberg
model on the pyrochlore lattice. These show that the apparent order of
the transition is indeed very sensitive to the exchange constants.

We have supplemented the numerics by a perturbative renormalization
group (RG) analysis. It is known that ``fluctuation-induced
first-order transitions'' occur when there is no stable fixed point in
a perturbative RG calculation. This frequently occurs when the number
of components of the order parameter $n$ is larger than
4,\cite{Brezin,Bak,Mukamel1,Mukamel2}. In general, such transitions
are expected to be only \textit{weakly} first-order. Another
microscopic scenario for a first order transition is the proposal of
Brazovskii\cite{Brazovskii} that the existence of a continuous set of
degenerate soft modes could change the order of the transition.  It
was latter shown that the RG analysis of models with soft modes along
special directions lacks stable fixed points and the models are indeed
likely to undergo first-order
transitions.\cite{Mukamelsmectic,Grinstein} This scenario is, for
instance, relevant to the description of the liquid crystal transition
where the degeneracy naturally comes from the isotropy of the
liquid.\cite{Mukamelsmectic} It is also particularly relevant in
frustrated magnets where frustration may, precisely, provide a large
number of quasi-degenerate soft-modes, though with different
geometrical structures. We shall consider a similar RG analysis,
applicable for the symmetry of the dipolar pyrochlores which have
lines of degeneracy, in this paper.

In this work, we study the order of the transition of the classical
Heisenberg model on the pyrochlore lattice with long-range
dipole-dipole interactions and exchange interactions. We show, by
means of Monte-Carlo simulations, that the transition is
\textit{strongly} first-order when a continuous set of soft modes is
present in $J({\bf q})$, and becomes \textit{weakly}
first-order when that degeneracy is removed by including further
neighbor interactions (Sec.~\ref{montecarlo}).  In section \ref{RG},
we present a Landau-Ginzburg-Wilson simplified model and a
perturbative renormalization group analysis of the transition, that
predicts a first-order transition, in agreement with the numerical
results in Sec.~\ref{montecarlo}. We summarize our conclusions in
Sec.~\ref{conclusion}.

\section{Monte-Carlo Simulations}
\label{montecarlo}

We present in this section some results of Monte Carlo simulations,
using the parallel tempering approach,\cite{pt} on the Heisenberg
model with dipolar and exchange interactions on the pyrochlore
lattice.

The classical
Heisenberg Hamiltonian is:
\begin{eqnarray} 
{\cal H} &=& \sum_{\langle i,j\rangle} J_{ij} \textbf{\mbox{S}}_i \cdot
\textbf{\mbox{S}}_j \nonumber \\ &+& (g \mu_B)^2 \sum_{\langle i,j\rangle}
\left(
\frac{\textbf{\mbox{S}}_i \cdot \textbf{\mbox{S}}_j}{r_{ij}^3} - 3
\frac{(\textbf{\mbox{r}}_{ij} \cdot \textbf{\mbox{S}}_i)
(\textbf{\mbox{r}}_{ij} \cdot \textbf{\mbox{S}}_j)}{r_{ij}^5} \right)
\end{eqnarray} 
where $ \textbf{\mbox{S}}_i$ is a classical spin vector of length
$S=7/2$ (for Gd$^{3+}$) on site $i$, and $J_{ij}$ is the Heisenberg
exchange between the neighbors: we will consider the first ($J$), second
($J_2$) and third neighbor ($J_3$) couplings.\cite{Cepas}

The number of spins is $N = 16L^3$ ($L\le 4$) and periodic boundary
conditions are applied. The factor of 16 arises because the pyrochlore
lattice consists of an fcc lattice of tetrahedra, each tetrahedron has
4 spins, and there are 4 sites of the fcc lattice in the conventional
cubic cell.  To incorporate the B-phase with periodic boundary
conditions, we need $L$ to be even, so most of our results are for
$L=2$ and 4.  Often the long-range dipolar interactions are cut-off
beyond a couple of neighbors\cite{PetrenkoGarnet} to speed up the
simulations.  However, here we have kept a large number of neighbors
(practically infinite) in order to reproduce accurately the structure
of the degenerate states. If the dipole-dipole interaction is cut-off,
ripples appear in the degenerate lines of soft
modes\cite{Raju,Cepas,Gingras}. We constructed periodic repetitions of the
Monte Carlo clusters and included the contributions of many blocks in
performing the dipolar sums. Because there is no cut-off in the dipole
interactions, the simulation becomes slow for large sizes, so we are
limited to $L \le 4$ ($N \le 1024$).

We investigate ordering at
${\bf q}=0$
(called A)
see Fig.~\ref{Astate}, and at
${\bf q}=\mbox{\boldmath$\pi$}$ (called B), see Ref. [\onlinecite{Cepas}].
With A ordering the order parameter has $n=3$ components and with B ordering
it has $n=4$ components as detailed in Sec.~\ref{RG}.
The $n$-component order parameters are given by
\begin{eqnarray}
\psi &=& (\psi_1,\dots,\psi_n) \\ 
\psi_\alpha &=& \frac{1}{N} \sum_{i=1}^N {\bf S}_i \cdot
{\bf e}_{i}^{(\alpha)}
\end{eqnarray}
where ${\bf e}_{i}^{(\alpha)}$ is the unit vector of spin $i$ assuming
the system is fully ordered in component $\alpha$ of ordering type A
with the $(xy)$ state (resp. $(xz)$, $(yz)$) corresponding to $\psi=(1,0,0)$
(resp. $\psi=(0,1,0)$, $\psi=(0,0,1)$) (see e.g.~the arrows in
Fig.~\ref{Astate}) or B.  In the course of the simulation of a finite
system, the spin configuration can fluctuate between different,
equivalent ordered states. We therefore compute the invariant
quantities
\begin{equation}
m^{(2)} =\sum_{\alpha=1}^n \langle \psi_\alpha^2 \rangle;
\hspace{0.6cm}
m^{(4)} = \langle \left(\sum_{\alpha=1}^n \psi_\alpha^2
\right)^2 \rangle,
\end{equation}
It is convenient to also compute the dimensionless Binder ratio
\begin{equation}
g = {1 \over 2} \left[ (n + 2) - n {m^{(4)} \over
\left(m^{(2)}\right)^2} \right] ,
\end{equation}
for both A and B orderings, 
which has the property that it tends to 0 at high temperature and to 1 in an
ordered state. (Remember that $n=3$ for A-type ordering and $n=4$ for B-type
ordering.)

\begin{figure}[htbp]
\centerline{ \psfig{file=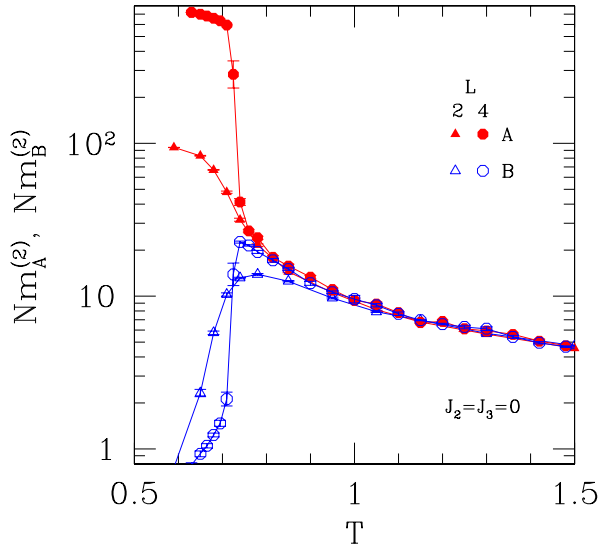,width=8cm,angle=-0}}
\vspace{-1cm}
\caption{(Color online). Order parameter squared ($\times N$)
for the A phase, shown in Fig.~\ref{Astate}, and the B phase,
for $J_2=J_3=0$ as
function of temperature $T$ in a Monte Carlo simulation of the dipolar
pyrochlore Heisenberg
antiferromagnet. We see a large jump in the order parameter
for the largest system size $L=4$
signaling the onset of a strong first-order transition. 
The number of spins is given by $N = 16L^3$.}
\label{montecarlosimulation1}
\end{figure}

\begin{figure}[htbp]
\centerline{ 
\psfig{file=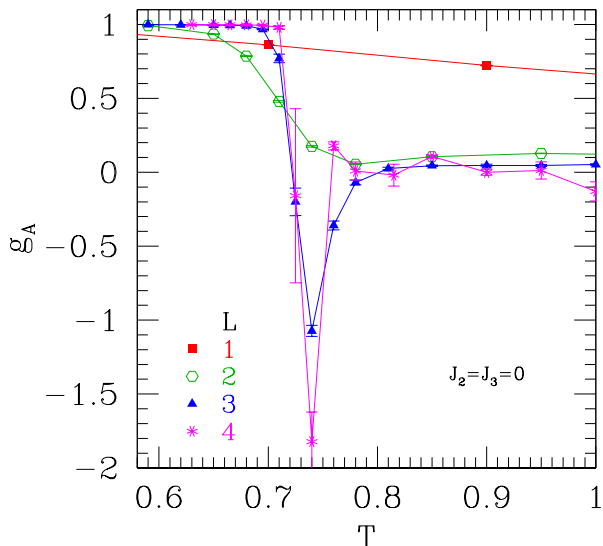,width=8cm,angle=-0} }
\vspace{-1cm}
\caption{(Color online). Binder ratio for
the A phase for different system sizes $L$ for $J_2 = J_3 = 0$.}
\label{montecarlosimulation2}
\end{figure}

Firstly we consider the dipolar model with only nearest neighbor interactions.
Results for the order parameters are shown in
Fig.~\ref{montecarlosimulation1}.
The parameter $J$ is taken to be the same as in Gd$_2$Ti$_2$O$_7$,
i.e. from the high temperature susceptibility $J=0.4 $K, and the
dipole-dipole interactions are fixed by the inter-ion
distances.\cite{Raju,Bonville} The order parameter associated with the A state
jumps rapidly at $T \simeq 0.7$ K, indicating a strong first order
transition at that temperature,
whereas that of the B state goes to zero at the same temperature.
These results
gives a much lower value of the transition temperature $T_N$ than the
mean-field theory value of $5.3$ K, though it is a bit smaller than the
experimental value of 1 K.  

The Binder ratio shown in Fig.~\ref{montecarlosimulation2} is also
strongly discontinuous.  In the vicinity of $T_c$ the Binder ratio
gets negative, as expected for a first-order phase
transition.\cite{Binder} Given the results of
Figs.~\ref{montecarlosimulation1} and \ref{montecarlosimulation2}, we
conclude that the transition is \textit{strongly} first-order for $J_2
= J_3 = 0$. The very small preference for the
B-phase\cite{huse:02,Cepas,Gingras} indicated by the small minimum of
$J(\mathbf{q})$ at $\mathbf{q} =\bpi$, which would be relevant at a
second order transition, is unimportant here because the transition is
so strongly first order.

\begin{figure}[htbp]
\centerline{ 
\psfig{file=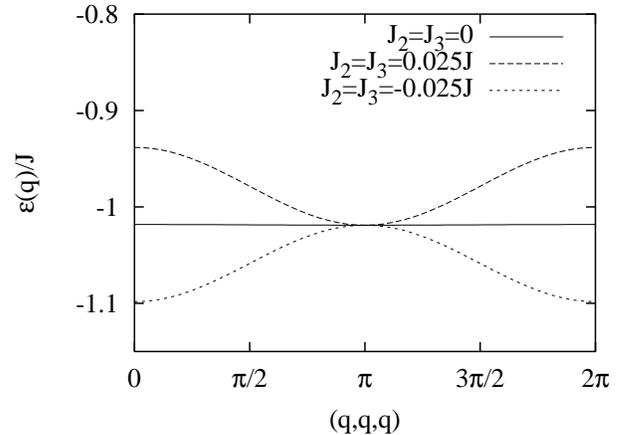,width=6cm,angle=-90} }
\caption{The lowest part of the spectrum of $J({\bf q})$ (from
Ref. [\onlinecite{Cepas}]). One sees that the degeneracy of the critical soft
modes is lifted by second and third neighbor couplings.
A positive (resp. negative)
$J_2=J_3$ favors ${\bf q}=\mbox{\boldmath$\pi$} $ (resp. ${\bf
q}=0$).}
\label{degeneracyJ2J3}
\end{figure}

We now include further neighbor interactions $J_2$ and $J_3$ that lift
the degeneracy of the dipolar model and select other
states, as studied in detail in Ref.~[\onlinecite{Cepas}]. The lowest part of
the spectrum of $J(\mathbf{q})$ is shown in
Fig. \ref{degeneracyJ2J3} for $\mathbf{q}$ along the $(1,1,1)$ direction. We
see that spectrum is almost precisely degenerate for $J_2 = J_3 = 0$ but that
$\mathbf{q}=0$ is preferred if $J_2$ and $J_3$ are negative, while
$\mathbf{q}=\mbox{\boldmath$\pi$}$ is preferred if $J_2$ and $J_3$ are
positive.  For simplicity we restrict ourselves to $J_2=J_3$ and study how the
character of the transition is modified relative to the case $J_2 = J_3 = 0$. 

\begin{figure}[htbp]
\centerline{ \psfig{file=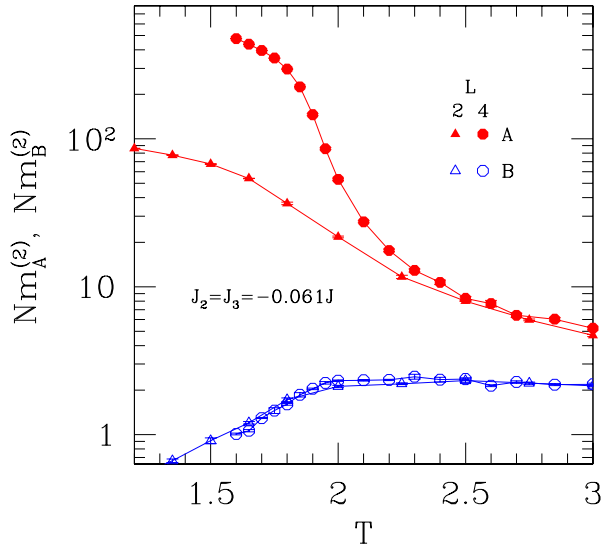,width=8cm,angle=-0}}
\vspace{-1cm}
\caption{(Color online). Order parameters squared ($\times N$) for
$J_2=J_3=-0.061J$.  A transition occurs to the A phase (${\bf q}=0$)
which is more gradual than for the degenerate case with $J_2=J_3=0$
shown in Fig.~\ref{montecarlosimulation1}.  }
\label{montecarlosimulationJ1J21}
\end{figure}

\begin{figure}[htbp]
\centerline{ 
\psfig{file=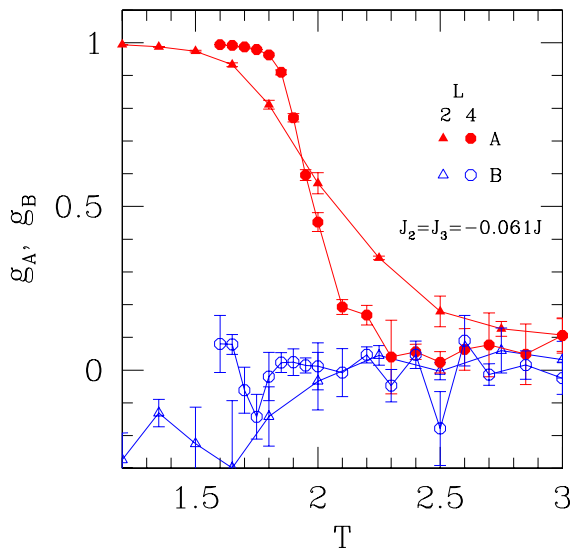,width=8cm,angle=-0} }
\vspace{-1cm}
\caption{(Color online). The Binder ratio for
the A and B phases for different system sizes $L$ for $J_2=J_3=-0.061J$.}
\label{montecarlosimulationJ1J22}
\end{figure}

For $J_2=J_3<0$, we find the same A state as for
$J_2=J_3=0$, but the
transition temperature shifts to higher temperatures, see
Figs.~\ref{montecarlosimulationJ1J21} and
\ref{montecarlosimulationJ1J22}. This is 
expected from
Fig.~\ref{degeneracyJ2J3} since,
with $J_2=J_3<0$,
$J({\bf q})$ acquires a
well-defined minimum at ${\bf q}=0$ which gets deeper with increasing $J_2$
and $J_3$.

Even with couplings as small as $J_2=J_3=-0.061J$, the order parameter
and the Binder ratio, shown in Fig.~\ref{montecarlosimulationJ1J22},
vary in a much more gradual way than for the degenerate case
$J_2=J_3=0$.  Although only finite-size scaling on a bigger range of
sizes could say whether the transition is first or second order, it is
clear that removing the degeneracy makes the transition less
first-order compared with the degenerate case.  These results are
consistent with earlier simulations on model without dipole-dipole
interactions (where larger clusters could be considered) which pointed
out a continuous\cite{Reimersexponents} or \textit{weakly}
first-order\cite{Mailhot} transition for the pyrochlore lattice with
$J_3<0$.

\begin{figure}[htbp]
\centerline{ \psfig{file=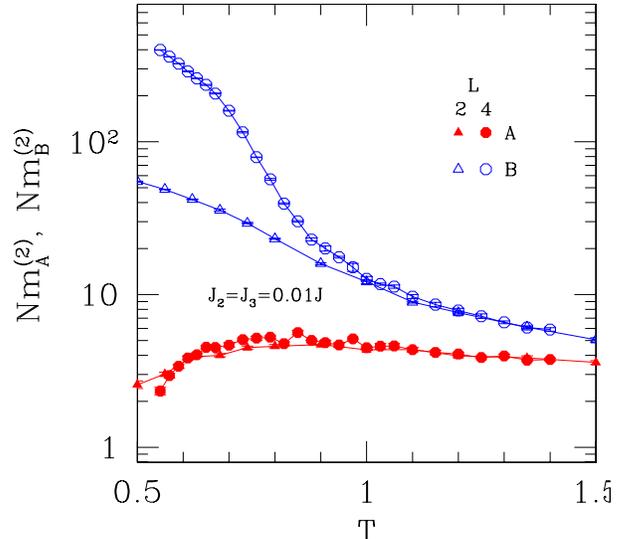,width=8cm,angle=-0}}
\vspace{-1cm}
\caption{(Color online). Order parameters obtained by Monte Carlo
simulation for $J_2=J_3=0.01J$.
The transition gives rise to the B-like state (${\bf
q}=\mbox{\boldmath$\pi$}$) and is more gradual than for the case of
$J_2=J_3=0$ shown in Fig.~\ref{montecarlosimulation1}.}
\label{montecarlosimulation3}
\end{figure}

\begin{figure}[htbp]
\centerline{ 
\psfig{file=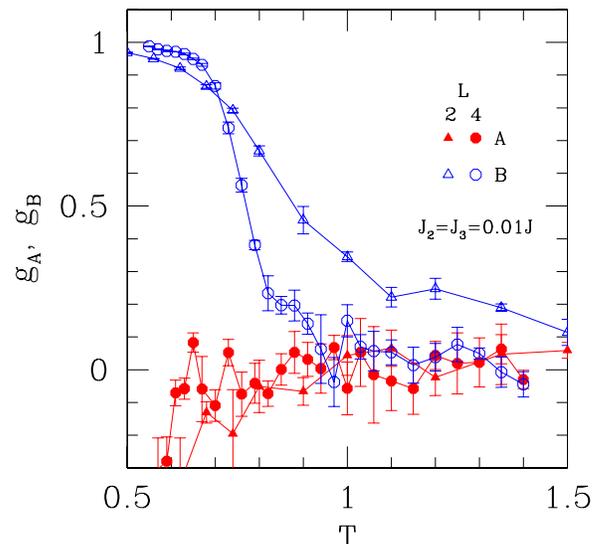,width=8cm,angle=-0} }
\vspace{-1cm}
\caption{(Color online). Binder ratio for
the A and B phases for different system sizes $L$ for $J_2=J_3=0.01J$.}
\label{montecarlosimulation4}
\end{figure}

Next we consider $J_2=J_3>0$ which, from Fig.~\ref{degeneracyJ2J3}, is
expected to favor the B-like states (${\bf q}=\mbox{\boldmath$\pi$}$),
and indeed this is the case as shown by
Figs.~\ref{montecarlosimulation3} and \ref{montecarlosimulation4}.
The minimum of $J(\mathbf{q})$ at ${\bf q}=\mbox{\boldmath$\pi$}$
shown in Fig.~\ref{degeneracyJ2J3} is independent of $J_2$ and $J_3$
which implies that the mean-field transition temperature is also
independent of $J_2$ and $J_3$.  Although we do not have enough system
sizes to attempt a serious estimate of $T_N$ using finite size
scaling, it seems that the transition temperature is indeed quite
similar to that for $J_2=J_3 = 0$.

As was also found for $J_2=J_3<0$,
the transition is much more gradual than for the degenerate case,
showing that removing the degeneracy reduces the first-order character
of the transition. The sensitive dependence of both the order of the
transition and the nature of the ordered phase on $J_2$ and $J_3$ points to
the possible relevance of these terms in explaining the difference between
Gd$_2$Sn$_2$O$_7$ and Gd$_2$Ti$_2$O$_7$.

While the data for $J_2=J_3=0$ clearly indicate a strong first order
transition, our results for $J_2$ and $J_3$ non-zero (with either sign) are
not conclusive as to the order of the transition. The smooth behavior of the
Binder ratios shown in Figs.~\ref{montecarlosimulationJ1J22} and
\ref{montecarlosimulation4} is typical for a second order
transition. However, we cannot rule out the possibility of a
\textit{weak} first order transition where the correlation length at the
critical point, $\xi_c$,
is large. In this case, if $L < \xi_c$ 
the behavior will look like that of a second order
transition. Only for sizes where $L > \xi_c$ can one see a crossover to
behavior expected at a first order transition. Hence for $J_2$ and $J_3$
non-zero, where the degeneracy is removed, the transition is either second
order or weakly first order. 
However, comparing Figs.~\ref{montecarlosimulationJ1J22}
and \ref{montecarlosimulation4} with the
corresponding figure for $J_2=J_3=0$, Fig.~\ref{montecarlosimulation2},
we see that behavior for $J_2$ and $J_3$ non-zero is very different from the
strong first order behavior found in the degenerate case.

\section{RG Analysis of the (1,1,1) model}
\label{RG}

We now study a Landau-Ginzburg-Wilson model by means of the
renormalization group analysis. Although the method is usually aimed
to study second-order phase transitions, the lack of stable fixed
points is often considered as an indication for a first-order kind of
transition.

\begin{figure}[htbp]
\centerline{
\psfig{file=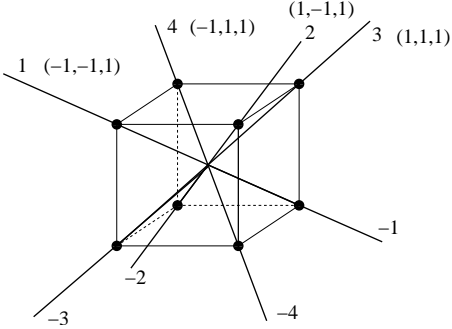,width=6cm,angle=-0}}
\caption{Lines of minimum energy in reciprocal space, given by the 4
equivalent (1,1,1) directions (the cube is drawn for convenience). The
(lack of) dispersion along the (1,1,1) lines is shown in
Fig. \ref{degeneracyJ2J3}.}
\label{degeneracies}
\end{figure}

Given the degeneracies of the soft modes with ${\bf q}$ along the 4
equivalent (1,1,1) directions when $J_2=J_3=0$ (see
Fig. \ref{degeneracyJ2J3}), the fluctuations of all these modes must
be taken into account simultaneously. For this reason, we consider a
model with an infinite-component order parameter (extended to
dimension $d$) and the fluctuations with wave-vectors close to these
soft mode directions. The quadratic part of the Hamiltonian is
written:
\begin{eqnarray}
H^{(2)} &=& \int \frac{d^d {\bf q}}{(2 \pi)^d} \sum_{i=1}^4 {\cal G}^{-1}_{0i}({\bf q}) \psi_{i {\bf q}} \bar{\psi}_{i {\bf q}} \\
 {\cal G}^{-1}_{0i}({\bf q}) &=& (r+ {\bf q}_{\perp, i}^2 + a q_{\parallel, i}^{2m}) \\
{\bf q}_{\perp,i} &=& {\bf q} - (\hat{v}_i.\hat{{\bf q}}) {\bf q} 
\end{eqnarray}
where the $\hat{v}_i$ are of norm 1 and represent the $i=1,...,4$
(1,1,1) directions given in Fig. \ref{degeneracies}.  We have 8
fields, $ \psi_{i {\bf q}}$ and $\bar{\psi}_{i {\bf q}}$
($i=1,\dots,4$), with $\bar{\psi}_{i {\bf q}}=\psi_{-i, -{\bf
q}}=\psi^{*}_{i {\bf q}}$. If we ignore the $ a q_{\parallel, i}^{2m}$ term,
then,
when $r=0$, all the modes with ${\bf
q}_{\perp,i}=0$ become simultaneously unstable. However, as in previous
works,\cite{Mukamelsmectic,Grinstein} we include the small dispersion
along the (1,1,1) lines, $a q_{\parallel, i}^{2m}$, to make the calculation
well defined at intermediate stages.  To study the degenerate case, it
will be eliminated at the end of the calculation by taking $m \rightarrow
+\infty$.

\noindent The fourth-order invariants are similar to that introduced to describe the nematic smectic-C transitions\cite{Mukamelsmectic} and are given by:

\begin{equation}
 H^{(4)} =  \int  \frac{d^d{\bf q}_1}{(2 \pi)^d} \frac{d^d{\bf q}_2}{(2 \pi)^d} \frac{d^d{\bf q}_3}{(2 \pi)^d}  \frac{d^d{\bf q}_4}{(2 \pi)^d}\delta_{{\bf q}_1,{\bf q}_2,{\bf q}_3,{\bf q}_4}  {\cal H}_4 
\end{equation}
where $\delta_{{\bf q}_1,{\bf q}_2,{\bf q}_3,{\bf q}_4}$ ensures that
the total momentum of the four $\psi$ is zero. For instance, we can
cancel the momentum by choosing pairs of momenta along the same
(1,1,1) line, \textit{i.e.} by combining $\psi_{i,{\bf q}_1}$ with
$\psi_{-i,{\bf q}_2=-{\bf q}_1}$, and similarly with ${\bf q}_3$ and
${\bf q}_4$. That gives a fourth-order term $u_p \psi_i \bar{\psi}_i
\psi_{i+p} \bar{\psi}_{i+p}$ term, where $i=1,\dots,4$ and $(i+p)$ is
meant for $(i+p-4)$ if $(i+p)>4$. Note that given the $C_4$ symmetry,
$u_p$ does not depend upon $i$, but only upon $p=0,1,2$.  In addition,
we could choose the first two ${\bf q}$'s along 1 and 3 for instance
(see Fig. \ref{degeneracies}) and the other two along -2 and -4, which
gives $ \psi_1 \bar{\psi}_2 \psi_3 \bar{\psi}_4$. Another simplifying
feature that we have adopted consists of neglecting the wave-vector
dependence of the coefficients $u_p$. Omitting to write the ${\bf
q_{1,2,3,4}}$ wave-vectors, the only fourth-order invariants are given
by:
\begin{eqnarray}
 {\cal H}_4 &=& \sum_{p=0}^2 u_p \sum_{i=1}^4 \psi_i \bar{\psi}_i \psi_{i+p} \bar{\psi}_{i+p}  \nonumber  \\ &+& \frac{1}{2} u_3  \left( \psi_1 \bar{\psi}_2 \psi_3 \bar{\psi}_4 + \bar{\psi}_1 \psi_2 \bar{\psi}_3 \psi_4 \right)
\end{eqnarray}
 where $\bar{\psi}_i=\psi_{-i}$. We call $H^{(2)}+H^{(4)}$ the (1,1,1)
 model. First, we consider the Hartree correction\cite{Brazovskii} to
 the gap $r$ (self-energy):
\begin{eqnarray}
{\cal G}_{i}^{-1} & \equiv & r + {\bf q}_{\perp i}^2 + \Sigma_{i}(r) \\
\Sigma_{i}(r) &=& \frac{1}{6}(3u_0 + 2u_1 +u_2) \int  \frac{d^d{\bf q}}{(2 \pi)^d} {\cal G}_{0i}({\bf q})
\label{Hartree}
\end{eqnarray}
In $d=3$, if we introduce momentum cut-offs, the new gap $r^{\prime}$ ($a \rightarrow 0$) is given by:
\begin{equation}
r^{\prime} = r +  \alpha \Lambda^{\prime} \ln(1+\frac{\Lambda^2}{r^{\prime}})
\end{equation} 
where $\alpha$ is a proportionality coefficient and $ \Lambda$ and $
\Lambda^{\prime}$ the cut-offs. Due to the strong singularity of the
right-hand-side, the gap does not vanish anymore. It suggests that the
paramagnetic phase remains locally stable below the transition, together with
other more stable phases. From the existence of other phases (at least
at the mean-field level), the transition is expected to be
first-order.\cite{Brazovskii}

However, the cut-offs $\Lambda$ and
$\Lambda^{\prime}$ enter explicitly the equation and a more controlled
result can be obtained by the renormalization group analysis by
restricting the integration to a shell of momentum
$\Lambda/b<q<\Lambda$ with $b>1$.\cite{Mukamelsmectic,Grinstein} For
this we introduce new real order-parameters ($i=1,...,4$) :
\begin{equation}
\psi_{i}= \phi_i + i \bar{\phi}_i \hspace{1cm} \bar{\psi}_{i}= \phi_i - i \bar{\phi}_i
\end{equation}
and the quadratic and quartic terms become:
\begin{eqnarray}
{\cal H} &=& \sum_{i=1}^4 {\cal G}^{-1}_{0i} ( \phi_i^2 +  \bar{\phi}_i^2 ) \label{H2}  +  u_0 \sum_{i=1}^4 (\phi_i^2 + \bar{\phi}_i^2)^2  \nonumber \\  &+& u_1 [(\phi_1^2 + \bar{\phi}_1^2)+(\phi_3^2 + \bar{\phi}_3^2)] [(\phi_2^2 + \bar{\phi}_2^2)+(\phi_4^2 + \bar{\phi}_4^2)] \nonumber \\ &+& u_2 [(\phi_1^2 + \bar{\phi}_1^2) (\phi_3^2 + \bar{\phi}_3^2) +  (\phi_2^2 + \bar{\phi}_2^2) (\phi_4^2 + \bar{\phi}_4^2)] \nonumber  \\
&+& u_3 ( \phi_1 \phi_2 \phi_3 \phi_4 +  \bar{\phi}_1  \bar{\phi}_2  \bar{\phi}_3  \bar{\phi}_4 ) \nonumber \\
&-& u_3  (\phi_1  \bar{\phi}_2  \phi_3 \bar{\phi}_4 +  \bar{\phi}_1 \phi_2 \bar{\phi}_3  \phi_4 ) \nonumber \\
&+& u_3 (\phi_1  \bar{\phi}_3+  \bar{\phi}_1 \phi_3)( \phi_2 \bar{\phi}_4+  \bar{\phi}_2 \phi_4) \label{H4}
\end{eqnarray} 
The derivation of the RG equations for the coupling constants is
then similar to that of Ref. [\onlinecite{Mukamelsmectic}] except that
we have to keep track of the field labels, given the anisotropy of ${\cal G}_{0i}$ in
Eq. (\ref{H2}). By integrating over a shell of momentum $\Lambda/b <q<
\Lambda$, we find the recursion relations for the new coupling constants:
\begin{eqnarray}
u_0^{\prime} &=& b^{\epsilon} \left[u_0 - \right( 40 u_0^2 I_0 + 4
u_1^2 I_0 + 2 u_2^2 I_0 \left) \right] \nonumber \\ u_1^{\prime} &=&
b^{\epsilon} \left[u_1 - \right( 8 u_1^2 I_1 + u_3^2 I_1 + 32 u_0 u_1
I_0 + 8 u_1 u_2 I_0 \left) \right] \nonumber \\ u_2^{\prime} &=&
b^{\epsilon} \left[u_2 - \right( 8 u_1^2 I_0 + 8 u_2^2 I_2 + u_3^2 I_2
+ 32 u_0u_2 I_0 \left) \right] \nonumber \\ u_3^{\prime} &=&
b^{\epsilon} \left[u_3 - \right( 8 u_2 u_3 I_2 + 8 u_1 u_3 I_1 \left)
\right] \nonumber \\ a^{\prime} &=& b^{-2(m-1)} a \label{dangerous}
\end{eqnarray}
where $\epsilon=4-d$ and the integrals are defined by
\begin{equation}
I_p = \int_{\Lambda/b}^{\Lambda} \frac{d^d {\bf q}}{(2 \pi)^d} {\cal G}_{0i}({\bf q}) {\cal G}_{0i+p}({\bf q}) 
\end{equation}

\subsection{Degenerate (1,1,1) model ($a \rightarrow 0$)}

The relation (\ref{dangerous}), together with the divergences of the
integrals $I_p$ for $a \rightarrow 0$ (at $r=0$, finite $b$) implies
that $a$ is a dangerous irrelevant variable for $m>1$. $I_0$ diverges
indeed as $a^{(d-5)/2}$, and $I_{1,2}$ as $a^{(d-3)/2}$. To take into
account these divergences, we have to introduce rescaled constants
$\tilde{u}_i= u_i a^{(d-5)/2} $
(Ref. [\onlinecite{Mukamelsmectic,Grinstein}] and references
therein). With $a^{(5-d)/2} I_{1,2} \rightarrow 0$, the recursion
relations become (we introduce ${\cal I} \equiv \mbox{lim}_{a
\rightarrow 0} a^{(5-d)/2} I_{0}$):
\begin{eqnarray}
\tilde{u}_0^{\prime} &=& b^{m(5-d-1/m)} \left[\tilde{u}_0 - \right( 40 \tilde{u}_0^2  + 4 \tilde{u}_1^2 + 2 \tilde{u}_2^2 \left) {\cal I} \right]  \nonumber \\
\tilde{u}_1^{\prime} &=& b^{m(5-d-1/m)} \left[\tilde{u}_1 - \right( 32 \tilde{u}_0 \tilde{u}_1 + 8 \tilde{u}_1 \tilde{u}_2 \left) {\cal I} \right]  \nonumber \\
\tilde{u}_2^{\prime} &=& b^{m(5-d-1/m)} \left[\tilde{u}_2 - \right( 8 \tilde{u}_1^2 + 32 \tilde{u}_0 \tilde{u}_2 \left) {\cal I} \right] \nonumber  \\
\tilde{u}_3^{\prime} &=& b^{m(5-d-1/m)} \tilde{u}_3  
\end{eqnarray}
All the fixed points are unstable for $d<5-1/m$ (the upper critical dimension is 5 for $m \rightarrow \infty$) since $\tilde{u}_3^{\prime} =
b^{m(5-d-1/m)} \tilde{u}_3$.  Although strictly speaking new fixed
points could occur at order $\epsilon^2$, the present calculation at
order $\epsilon$ is compatible with the first-order transition
observed in the Monte-Carlo simulations.

\subsection{No Degeneracy}

We now remove the degeneracy, \textit{e.g.} by including further neighbor
couplings in the microscopic Hamiltonian.

\subsubsection{$\mathbf{q = 0}$}
First of all we assume that the ordering is at $\mathbf{q} = 0$. It is
simplest to go back to the order parameter shown in Fig.~\ref{Astate}, realize
that there are three components $\psi_i$, $i=1,2,3$,
and that the symmetry is cubic.
The Hamiltonian of this cubic model is therefore given by
\begin{equation}
{\cal H} =  \sum_{i=1}^3 \left[(r+ {\bf q}^2) \psi_{i}^2 +
u_0  \psi_i^4 \right] 
+ u_1 \sum_{i,j}  \psi_i^2  \psi_j^2 \, .
\end{equation} 
There has been a controversy regarding whether the stable
fixed point of the cubic model is the Heisenberg or the cubic fixed
point. A recent 6-loop expansion has shown that for $n>2.89$, the
stable fixed-point is the cubic one.\cite{Carmona} Depending on the
initial values for the coupling constants, the transition could be
either first-order or continuous. However, to stabilize the \textit{collinear}
states (with $\psi$ either $(1,0,0)$, $(0,1,0)$, or $(0,0,1)$
and ${\bf q}=0$), the set of initial coupling constants leads to
a first-order transition.\cite{Amit}

We have seen that a first order transition is obtained both in the
degenerate case and also when there is a well-defined minimum at ${\bf
q}=0$. Is, then, the degeneracy of soft modes important or not?  We
note that, in the absence of degeneracy, the transition may be only
\textit{weakly} first-order.  The problem was studied some years ago
in the context of the pyrochlore FeF$_3$. For this compound, the ${\bf
q}=0$ state found by neutron scattering was characterized by a
3-component order parameter,\cite{Reimersexponents} similar to the one
we have here. Monte-Carlo simulations have shown that the collinear
structures with $\psi=(1,0,0);(0,1,0);(0,0,1)$ are preferred,
but the transition first appeared to be second-order, with unusual
critical exponents\cite{Reimersexponents} contrary to the RG argument
given above. We can reconcile these results by suggesting that the
transition may be \textit{weakly} first-order, so that the correlation
length would exceed the size of the Monte-Carlo cluster and the
transition would appear second-order in the simulation. This is also
comforted by a reexamination of the Monte-Carlo results, which
suggested that the transition is more likely to be indeed weakly
first-order.\cite{Mailhot}

\subsubsection{$\mathbf{q = }\bpi$}
We now assume that the degeneracy is lifted in such a way that one of
the four $\mbox{\boldmath$\pi$}$ wave-vectors is selected. Since
$\mbox{\boldmath$\pi$}$ and $-\mbox{\boldmath$\pi$}$ are related by a
reciprocal lattice vector, we have to take into account the
fluctuations of four fields only, with ${\bf q}$ close to any of the
$\mbox{\boldmath$\pi$}$ wave-vectors,
$\psi_i,i=1,\dots,4$. The critical model is given by
\begin{eqnarray}
{\cal H} &=&  \sum_{i=1}^4 \left[(r+ {\bf q}^2) \psi_{i}^2 +  u_0  \psi_i^4  \right] + u_1 \sum_{i,j}  \psi_i^2  \psi_j^2  \nonumber \\
 &+& u_3  \psi_1 \psi_2 \psi_3 \psi_4 \, .
\end{eqnarray} 
This model is known to possess unstable fixed points at order
$\epsilon^2$.\cite{Mukamel2} Therefore the transition to the
$\mbox{\boldmath$\pi$}$ phases is also expected to be first-order.

On the basis of the LGW models alone, we would conclude that the phase
transitions in the dipolar pyrochlore are \textit{all} first-order in
character. Such a simple analysis does not say whether the transition
is \textit{strongly} or \textit{weakly} first-order that is quite a
relevant question when one comes to compare with
experiments. Nonetheless, the results presented in this section are
compatible with the Monte-Carlo simulations of section
\ref{montecarlo}. The latter are important, precisely to say whether
the transitions are weakly or strongly first-order.
 
\section{Conclusion}

\label{conclusion}
We have considered the dipolar Heisenberg model on a pyrochlore lattice with
nearest neighbor interactions and a small amount of second and third neighbor
interactions ($J_2$ and $J_3$). For $J_2=J_3=0$ the system is highly
degenerate, see Fig.~\ref{degeneracyJ2J3},
and fluctuation effects pick out ordering at
$\mathbf{q}=0$ (A-type). Monte Carlo simulations show that the transition is
very strongly first
order in this case,
in contrast to mean-field theory which predicts a second order
transition. A first order transition is also predicted by a renormalization
group analysis. When the degeneracy is removed by including $J_2$ and $J_3$ the
transition is more gradual, showing that the degeneracy is necessary to get a
\textit{strong} first order transition.
Given the limited range of sizes in the Monte
Carlo simulations, we cannot say from the simulations
whether the transition is second
order or weakly first order for $J_2=J_3 \ne 0$.
However, according to a renormalization group analysis
for the non-degenerate case, both A and B type orderings have
no stable fixed points, indicating, presumably, a fluctuation induced first
order transition. Usually this type of transition is only weakly first order,
and this seems to be consistent with our numerical data.

Because of the degeneracy for $J_2=J_3=0$, a small amount of second
and third neighbor coupling can also change the nature of the ground
state. We find that for $J_2=J_3< 0$ the A phase is retained but for
$J_2=J_3>0$ we obtain a $\mathbf{q}=\mbox{\boldmath$\pi$}$ (B-type)
ordering. In future work we will study in more detail the nature of
this B-type phase, and also consider other possible phases that occur
when $J_2 \ne J_3$. It is possible that anisotropic interactions,
suggested on the basis of high-temperature ESR\cite{Hassan} and by EPR
on diluted samples,\cite{Glazkov,note} may be needed to explain the
experimentally observed phases in detail.

Our results provide a natural explanation for Gd$_2$Sn$_2$O$_7$ having
a strong first-order transition,\cite{Bonville} while
Gd$_2$Ti$_2$O$_7$ has a second-order transition\cite{Champion} (though
a weak first-order transition is not ruled out experimentally); namely
second and third neighbor interactions are very weak in
Gd$_2$Sn$_2$O$_7$, but they are stronger and positive for
Gd$_2$Ti$_2$O$_7$. In this respect, \textit{ab-initio} calculations
could give some estimate of the strength of the couplings. This
picture is also consistent with the observations that
Gd$_2$Ti$_2$O$_7$ orders\cite{Champion,Stewart} at
$\mathbf{q}=\mbox{\boldmath$\pi$}$ while Gd$_2$Sn$_2$O$_7$ should be
A-type with equivalent sites and moments perpendicular to the local
(1,1,1) directions.\cite{Bertin}

In the presence of a magnetic-field,\cite{Ramirez,Cepas}
Gd$_2$Ti$_2$O$_7$ has a rich phase diagram. For the future, it would
also be interesting to perform a study of field-induced transitions in
Gd$_2$Sn$_2$O$_7$, since this starts off with a quite different state
in zero field. In addition to the multiple phase transitions expected
on the basis of mean-field theory, the field reduces the fluctuations
and, hence should reduce the strong first-order character of the
transition.

\acknowledgments

We would like to thank A.~Wills for sharing with us unpublished
results concerning Gd$_2$Sn$_2$O$_7$.  O.C. would like to thank
especially G.~Jackeli and T.~Ziman for very stimulating discussions
and also S.~Bramwell, V.~Glazkov, E.~Kats, L.~L\'evy, Y.~Motome,
R.~Stewart, and M.~Zhitomirsky.  A.P.Y.~acknowledges support from the
National Science Foundation under Grant No.~DMR
0337049. B.S.S.~acknowledges support from the National Science
Foundation under Grant No.~DMR 0408247. A.P.Y.~would also like to
thank the Institut Laue Langevin, where this work was started, for its
hospitality.


\begin{thebibliography}{99}
\bibitem{notedeg}
Strictly speaking the degeneracy is only approximate. It
is lifted in favor of $\mathbf{q} = (\pi,\pi,\pi)$ by a very small
amount.\cite{huse:02,Cepas,Gingras}
\bibitem{note} We have checked that a sizeable on-site easy-plane anisotropy as found in ref. [\onlinecite{Glazkov}] does not change the degeneracy of the soft modes.
\end{thebibliography}
\end{document}